\documentclass[a4paper]{ifacconf}

\usepackage{graphicx,amsmath,url}      % include this line if your document contains figures
\usepackage[round]{natbib}             % required for bibliography
\def\portugues{1} 
\ifx\portugues\undefined
\def\portugues{0}
\fi

\if\portugues0
   \usepackage[english]{babel}
  \else
   \usepackage[spanish,brazil,english]{babel}
\fi

\usepackage{icomma}
\usepackage{color}
\usepackage{xcolor}

\usepackage[T1]{fontenc}

\usepackage[utf8]{inputenc}

\usepackage{ae}

\if\portugues1
 { }
 { }
 { }
 
\fi

\begin{document}

\if\portugues1

% =====================================================================
% =====================================================================
% USE THIS PART IF THE TEXT IS IN PORTUGUES OR SPANISH
% =====================================================================
% If the manuscript is in Spanish, please change the texts adequately.
% =====================================================================
% 
\selectlanguage{brazil}
	
\begin{frontmatter}
%ON THE RELIABILITY OF COMPUTATIONAL CHAOS-BASED CRYPTOGRAHY FOR SECURE INFORMATION EXCHANGE
\title{On the reliability of computational chaos-based cryptography for information exchange}

\thanks[footnoteinfo]{This study was financed in part by the Coordenação de Aperfeiçoamento de Pessoal de Nível Superior - Brasil (CAPES) - Finance Code 001, the Ministry of Science, Technology, Innovation and Communications (MCTI) and the Federal University of São João del Rei.}

\author[First]{Santos, T. A.} 
\author[Second]{Magalhães, E. P.} 
\author[Third]{Fiorio, D. R.}
\author[Fourth]{Nepomuceno, E. G.}

\address{Grupo de Controle e Modelagem (GCOM)\\
UFSJ - Universidade Federal de São João del-Rei\\ Pça. Frei Orlando, 170 - Centro - 36307-352 - São João del-Rei, Minas Gerais, Brasil}

\address[First]{e-mail: contato@tsantos.com.br.}
\address[Second]{e-mail: eduardopintomagalhaes@gmail.com}
\address[Third]{e-mail: dyuliafiorio@hotmail.com}
\address[Fourth]{e-mail: nepomuceno@ufsj.edu.br}

%%%%%%%%%%%%%%%%%%%%%%%%%%%

\selectlanguage{english}
\renewcommand{\abstractname}{{\bf Abstract:~}}
\begin{abstract}                % Abstract of not more than 250 words.
This paper investigates the use of dynamical chaotic systems to encrypt and exchange images between different devices. Two devices were used to simulate the Cubic Map, having the same set of initial conditions, to generate an encryption key. Although both devices are floating-point compliant, the simulations, and consequently the encryption key, turned out to differ from one another. This indicates that many existing chaos-based encryption schemes are just special cases of computational arithmetic properties, in which some characteristics in the construction of the devices coincided. A method to mitigate such flaw was also presented.

\vskip 1mm% não altere esse espaçamento
\selectlanguage{brazil}
{\noindent \bf Resumo}:  
Este trabalho investiga o uso de sistemas caóticos dinâmicos para criptografar e trocar imagens entre diferentes dispositivos. Dois dispositivos foram usados para simular o mapa cúbico, com o mesmo conjunto de condições iniciais e gerar uma chave de criptografia. Embora ambos os dispositivos sejam compatíveis com a norma de ponto flutuante, as simulações e, consequentemente, a chave de criptografia, mostraram-se diferentes umas das outras. Isso indica que muitos esquemas existentes de criptografia baseados em caos são apenas casos especiais de propriedades aritméticas computacionais, nas quais algumas características na construção dos dispositivos coincidem. Um método para mitigar essa falha também foi apresentado. 
\end{abstract}

\selectlanguage{english}


%\selectlanguage{english}

\end{frontmatter}
\else
% ===============================================================
% ===============================================================
% USE THIS PART IF THE TEXT IS IN ENGLISH
% ===============================================================
% ===============================================================
% 

\begin{frontmatter}
%ON THE RELIABILITY OF COMPUTATIONAL CHAOS-BASED CRYPTOGRAHY FOR SECURE INFORMATION EXCHANGE
\title{On the reliability of computational chaos-based cryptography to securely exchange information between different devices}

\thanks[footnoteinfo]{The authors are grateful to the MEC/SISU University Extension Program (ProExt), the Ministry of Science, Technology, Innovation and Communications (MCTI) and the Federal University of São João-del Rei.}

\author[First]{Santos, T. A.} 
\author[Second]{Magalhães, E. P.} 
\author[Third]{Fiorio, D. R.}
\author[Fourth]{Nepomuceno, E.G.}

\address{Grupo de Controle e Modelagem (GCOM)\\
UFSJ - Universidade Federal de São João del-Rei\\ Pça. Frei Orlando, 170 - Centro - 36307-352 - São João del-Rei, Minas Gerais, Brasil}

\address[First]{e-mail: contato@tsantos.com.br.}
\address[Second]{e-mail: eduardopintomagalhaes@gmail.com}
\address[Third]{e-mail: dyuliafiorio@hotmail.com}
   \address[Fourth]{e-mail: nepomuceno@ufsj.edu.br}
\renewcommand{\abstractname}{{\bf Abstract:~}}   
   
\begin{abstract}                % Abstract of not more than 250 words.
This paper investigates the use of dynamical chaotic systems to encrypt images and exchange it between different devices. Two devices were used to simulate the Cubic Map, having the same set of initial conditions, to generate an encryption key. Although both devices are floating-point compliant, the simulations, and consequently the encryption keys, turned out to be a mismatch. This indicates that many existing chaos-based encryption schemes are just special cases of computational arithmetic properties, in which some characteristics in the construction of the devices coincided.
\end{abstract}


\end{frontmatter}
\fi

%==============================================================================

\section{Introduction}
\label{sec:Introducao}

\par
\hspace{0.3cm} 
Chaotic systems are non-linear deterministic systems that behave in a dynamical and complex way without time, depending on their initial conditions \citep{may1976simple}. Chaos-based cryptography method uses a chaotic set of equations to produce a pseudo-random encryption key. Such a feature can be useful for file exchanging, as it keeps in-between parties unable to obtain the information on transit. As the security and reliability of this procedure depend on obtaining the same key from both ends, concerns about the finite precision of computational arithmetic must emerge.

%paragraph to introduce and describe the development of chaotic systems and its simulation

The study of dynamical systems dates back from 384–322 BC, when Aristotle proposed laws for object motion, persisting all the way to Rayleigh-Bénard's convection model and the wide famous butterfly effect by E.N. Lorenz  \citep{monteiro}. Chaotic systems are dynamical systems that have a sensitivity to their initial conditions, are deterministic and non-periodic \citep{lorenz1963deterministic}. 

%bibliographic revision paragraph about chaos cryptography development

One way to encode files is to use a single pseudo-random sequence obtained from a chaotic map and there has been extensive research regarding more efficient methods. For example, \citet{Liu2018} proposes a one-dimensional chaotic map encoding scheme that is more attack-tolerant and \citet{HUA201580} uses a combination of sine and logistic maps to tackle the same problem and recently there have been many works regarding new cryptography schemes \citep{silvarealizaccao,CHAI2017197,Li2017,7932843,pop00001,pop00002,pop00003}.

%paragraph that shows non-repeatability of results across different devices

Chaos-based cryptography often relies on computer simulation of dynamical systems to generate a key. It is well known that the inherit finite precision of computers don`t yield exact results \citep{Goldberg:1991:CSK:103162.103163,Monniaux:2008:PVF:1353445.1353446,852391}. It is also known that the IEEE 754 standard doesn't specify a default rounding mode, leading to different results of the same operation across different software \citep{PEIXOTO201836}.
\par

When cryptography is used to ensure data security, it is important that this data remains unchanged when passed from one device to another. Therefore, when simulating the same system across devices, the results must be a match. However, when simulating the Lorenz's system in two different devices it was noticed that just the first element of the pseudo-orbits was corresponding, which corresponds to its initial condition. To solve this problem and increase the correspondences, it was proposed the application of the step function, in order to replace the elements of the pseudo-orbit by binary numbers.

\par
In this study it was shown that encrypted data exchange can fail to be interpreted if some aspects of computational arithmetic and finite computer precision were not taken into account. As any chaotic system, any variation of the initial conditions and subsequent iterations can lead to completely different results. The authors also demonstrated that it is possible to interpret the information if such characteristics are considered.

\section{Preliminary Concepts}
\label{sec:refteo}

This section presents the fundamental theoretical concepts for the accomplishment of this work.

\subsection{Chaotic dynamical systems}

Chaotic dynamical systems have been studied from the work of \citet{lorenz1963deterministic}. The accepted definition of chaos found in the literature can be properly explained by  \citet{banks1992devaney}.
\\

\noindent\textbf{Definition 1:} \textit{\citep{banks1992devaney}}. \textit{Let \(f: X\rightarrow X\) be a chaotic system. This system is chaotic when it has the three following properties:}
\\

\noindent1) \textit{f} is transitive;
\\
2) the periodic orbits of \textit{f}  are dense in \textit{X};
\\
3) \textit{f} is sensitive to the initial conditions.
\vspace{0.4cm}

As for the first condition, for any subset of \textit{U} end \textit{V} in \textit{X}, there is a positive integer number \textit{k}, where \(f^{k}(U) \cap V\) is a is a non-empty set. The second condition refers to a density of periodic orbits of \textit{f} in \textit{X}.

%Sobre a primeira condição, para qualquer subconjunto de \textit{U} e \textit{V} em \textit{X}, existe um número inteiro positivo \textit{k}, tal que \(f^{k}(U) \cap V\) é um conjunto não vazio. A segunda condição refere-se a densidade de órbitas periódicas de \textit{f} em \textit{X}. % Diz-se que \textit{V} é denso em \textit{U} se para qualquer ponto \(u \in  U\) e para qualquer \(\varepsilon  >  0\) há um ponto

\textit{v} \(\varepsilon\) \textit{V}, such that \(\left \| u - v \right \| > \epsilon\). The third condition, \textit{f} is sensitive to initial conditions if there is a positive real number \(\delta   >  0\), such that for any point \(x \in  X\) and all neighborhood \textit{N} of \textit{x} exists a point \(y \in  N\) and \(n \in  
Set of Natural Numbers\), such that \(d(f^{n}(x),f^{n}(y)) >\epsilon \) where \textit{d} is the distance in \textit{X}.

For the system to be chaotic the system constants mus be: $\rho = 24.74$, $\sigma = 10$ e $\beta = 8/3$. (Jambersi) (Andrucioli, 2008).

\subsection{Orbits and pseudo-orbits and Lower Bound Error}

\subsubsection*{Orbits and pseudo-orbits} 

\textbf{Definition 1:} \textit{An orbit is a sequence of values from a map that can be represented by:}
%Uma órbita é uma sequência de valores de uma mapa que pode ser representada por:

\begin{equation}
    \left \{ x_{n} \right \} = \left [ x_{1},x_{2},x_{3},...,x_{n} \right ]
\end{equation}

\noindent\textbf{Definition 2:} \textit{A pseudo-orbit is an approximation of an orbit that can be represented by:}
%Uma pseudo-órbita é uma aproximação de uma órbita que pode ser representada por:

\begin{equation}
\hat{x}_{i,n} = {\hat{x}_{i,0}}, \hat{x}_{i,1}, \hat{x}_{i,2},...,\hat{x}_{i,n} 
\end{equation}
\\ such that

\begin{equation}
\left | x_{n} - {\hat{x}_{i,n}}  \right |\leq {\delta _{i,n}} 
\end{equation}

where the \(\delta _{i,n} \in\) real numbers  is the error \(\delta _{i,n}\geq 0\).

%\subsection{Extensão Intervalar}

%\textbf{Definition 1:} \textit{(Extensão Intervalar \citep{moore1979methods}). Seja f uma função da variável real x. Uma extensão
%intervalar de f é uma função de intervalo F de uma variável de intervalo X com a propriedade}

%\begin{center}
%\textit{F(x) = f(x)}
%\end{center}

%\textit{em que um intervalo é dado por um conjunto fechado de números reais x} $\epsilon \mathbb{R}$ tal que \(X = \left [ \underline{X}, \overline{X} \right ] = \left \{ x: \underline{X} \leqslant x \leqslant \overline{X} \right \}\)

%\noindent\textbf{Definition 2:} \textit{O e P são extensões intervalares equivalentes se}

%\begin{center}
%\textit{O(x) = P(x)}
%\end{center}

\subsubsection{Lower Bound Error}

The LBE (Lower Bound Error) is a tool used to obtain the lower limit of the computational error obtained at each iteration of computational simulation. It was developed by \citet{nepomuceno2017analysis} and its mathematical representation is given by:

\begin{equation}
    \delta _{a,n} = \left | \hat{x}_{a,n} - \hat{x}_{b,n}  \right |
\end{equation}

Where the pseudo-orbit is represented by $ \hat{x} $ and \textit{a} and \textit{b} are its indices indicating extensions.

\subsection{XOR Operation}

XOR encryption is an encryption encoding that is well known for its simplicity in modern encryption. XOR encoding is a symmetric encryption algorithm. As it is based on Boolean algebra, its principle is the derivative of the XOR function which returns "true" when two arguments have different values. The key is the same length as the message to be encrypted. And if the content of any message can be guessed or known, the key may be revealed. Within the context of encryption, the strength of the XOR cipher depends on the size and nature of the key. XOR encoding with a long random key may achieve better security performance \citep{xor}. A property of the XOR cipher is its statistical property that, for a perfectly random key stream, each encrypted bit has a 50 \% chance of being 0 or 1.

%XOR is an additive figure and it operates according to the following principle:

%\begin{subequations}
%\begin{equation}
%X \oplus 0 = X 
%\end{equation}

%\begin{equation}
%X \oplus  X = 0    
%\end{equation}

%\begin{equation}
%(X  \oplus  Y)  \oplus  Z = X \oplus  (Y \oplus  Z)    
%\end{equation}

%\begin{equation}
%(Y  \oplus  X)  \oplus  X = Y \oplus  (0 \oplus  Y) 
%\end{equation}

%\end{subequations}

%where $\oplus$ denotes the exclusive disjunction operation.

To exemplify the operation XOR encryption method, we have the following example:

The "SBAI" string (01110011-01100010-01100001-01101001  in 8-bit ASCII) can be encrypted with an arbitrary 01001100-01001100-01001100-01001100 key as follows:

\vspace{0.3cm}

\begin{equation}
\begin{split}
    \oplus  \frac{01110011-01100010-01100001-01101001}{01001100-01001100-01001100 -01001100} \\ \\ \hline \\ 
   = 01010011-01000010-01000001-01001001
\end{split}
\end{equation}
And decrypted as follows: \vspace{0.3cm}

\begin{equation}
\begin{split}
    \oplus  \frac{01010011-01000010-01000001-01001001}{01001100-01001100-01001100 -01001100} \\ \\ \hline \\ 
   = 01110011-01100010-01100001-01101001
\end{split}
\end{equation}

\subsection{Cubic Map}

The Cubic Map is a map that has a chaotic behavior from the value \textit{r}, known as the bifurcation parameter. This map is a discrete and dynamical system characterized by the equation below:

\begin{equation}
f_{r}(x) = rx^{3} + (1 - r)x
\end{equation}

\subsection{Shannon Entropy}

Shannon entropy $H$ is given by the formula 
\begin{equation}
H(x) = -\sum_{i=0}^{N-1} P_{i} log_{2}(P_i)
\end{equation}
where $P_i$ is the probability of number $i$ appearing in any position of the message. The normalized entropy output values from $0$ to $1$, where the higher the number, the higher the entropy and it can be given by

\begin{equation}
    H_{norm}=-\frac{H*log_2(2)}{log_2(length(p))}
\end{equation}

%%%%%%%%%%%%%%%%%%%%%%%%%%%%%%%%%%%%%%%%%%%%% Metodologia %%%%%%%%%%%%%%%%%%%%%%%%%%%%%%%%%%%%%%%%%%%%%%
\section{Methodology}
\label{sec:metodologia}
The current investigation involved sampling and comparing the results from the cubic map simulations on different devices. The system was selected based on the frequency it appears in the literature about chaos-based cryptography \citep{Rogers1983}.

Two devices, described in Table \ref{tb:devices} using GNU Octave 4.2.1 were used to compute the Cubic Map. The simulations had its initial condition set to $x_0=0.1$ and $r=3.6$ and were made on IEEE 754\citep{1985--ieee754} compliant environment to prevent different representation standards to interfere with results. The simulations ran until reaching $70*10^3$ iterations. This method was chosen because it is widely used by works regarding image cryptography.

%\caption{Describes each device used to simulate the chaotic system}
\begin{table}[!ht]
\centering
\caption{\label{tb:devices} Describes each device used to simulate the chaotic system.}
\begin{tabular}{lc}
\hline
Device & Description \\ \hline
1 & Intel(R) Core(TM) i7-7700HQ CPU @ 2.80GHz \\
2 & Intel(R) Xeon(R) CPU E5-2620 v4 @ 2.10GHz\\
\hline
\end{tabular}
\end{table}

The two simulations used the same interval extension to prevent error propagation, as previous studies shown that even using an equivalent expression to describe the same equation, if one changes the sequence of mathematical operations, results could  diverge from each other \citep{nepomuceno2016lower}.

The encryption process was based on the XOR operation. Figure \ref{fig:processnormal}(a) was selected as it is present in great part of the works regarding image cryptography. The image is represented by a 256x256 matrix of pixels named $Image$, each position being the intensity of the pixel expressed between a minimum and a maximum. This range is generally represented from 0 (0\%) (black) and 255 (100\%) (white). Thus, we need to manipulate the simulation results to obtain a key that has the same format as the image.

In order to obtain a key that matches the image pixel value range, the pseudo orbits were manipulated. For the encryption process, the pseudo-orbit obtained from Device 1 was used.  Each iteration was divided by 2, added 1, the first 3 decimal places were discarded, and the remainder were normalized as shown in Equation \ref{eq:norm}, where floor is defined as rounding towards $-\infty$. Such operations made possible to transform the pseudo-orbit from values in [-1,1] range into [0,255] range, as required to perform the cryptography.
 \begin{equation}
K_{norm}=floor(255*K)
 \label{eq:norm}
 \end{equation}
 
 It was also necessary to compose a 256x256 matrix consisting of the normalized results $K_{norm}$, that would be the key for the cryptography process. A matrix named $Key$ will be filled by $K_{norm}$ terms from top to bottom, left to right, until all positions are defined. One should note the importance of specify the exact order used to fill the matrix, as a different method could yield an unreadable decryption. 
 
 Finally, for the encryption phase, the XOR operation was used. The encryption process was made as in Equation \ref{eq:encryption}, using octave's \textbf{bitxor()} function. One should expect the original image if the first result is operated again with the same $Key$, which is thereby called decryption phase.
 \begin{equation}
 Key\oplus Image = EncryptedImage 
 \label{eq:encryption}
 \end{equation}

After obtaining the result, $EncryptedImage$ was sent through internet to Device 2. As for the decryption phase,  using the same set of initial conditions, Device 2 synthesized the same interval extension, went through the same manipulation and normalization techniques and compose it's own matrix, thereby called $Key2$. If $Key=Key2$ one should expect a reversible operation as shown in Equation \ref{eq:decryption}.
\begin{equation}
  EncryptedImage\oplus Key2 = Image
  \label{eq:decryption}
 \end{equation}

Considering the suspicions of error propagation due to floating point representation between different devices, which would lead to $Key\neq Key2$, a technique to reduce the largest Lyapunov exponent was applied. In this method, each iteration of the chaotic map was multiplied by $0.89$ before it was fed back into the equation, as this value was found to yield a satisfactory output. Such operation should reduce the minimal error and the largest Lyapunov exponent, also decreasing the lower bound error curve's slope \citep{llenepomuceno}. It shoud be noted that this technique only applies to small values of Lyapunov exponent, making necessary to develop new strategies to handle bigger values.

Also, a set of initial conditions was generated by obtaining 70 partitions of the interval (0,1) linearly. Each one of the partitions were an input to the chaotic map and simulated until reaching $1024$ iterations that filled a vector named $K$, using $r=3.61$.  Care was taken to reduce the simulation length for the same initial condition, to avoid ergodicity and have more accuracy throughout iterations. In order to check this results, an additional interval extension was simulated solely for this purpose. Aside from this multiplication, this method follows the exact same path as the above for both encryption and decryption phase.

For the sake of validating the results, the entropy of the images was first studied qualitatively by means of the calculation of Shannon Entropy and histogram comparison. The former is a standard test used in the literature about chaos simulations to validate outputs and the latter facilitate the visualization of the quantity of pixels within the same value at an image. It was also analyzed how both pseudo-orbits diverges as they were iterated, using the lower bound error. The logarithmic scale indicates error propagation as the difference between orbits grow. To validate the hypothesis that the output was chaotic, the largest Lyapunov exponent was obtained from the coefficient of the line formed by the linear regression of the lower bound error plot \citep{llenepomuceno}.

%Para a distinção da correlação entre as imagens original e criptografada, mostrando cada probabilidade de nível de cinza, foi utilizada a métrica histograma. Se a diferença do nível de cinza entre as imagens original e criptografada for grande, as imagens são não correlacionadas.

%%%%%%%%%%%%%%%%%%%%%%%%%%%%%%%%%%%%%%%%% RESULTADOS %%%%%%%%%%%%%%%%%%%%%%%%%%%%%%%%%%%%%%%%%%%%%%%%%
\section{Results}

\begin{figure}[ht!]
\centering
\includegraphics[width=8cm]{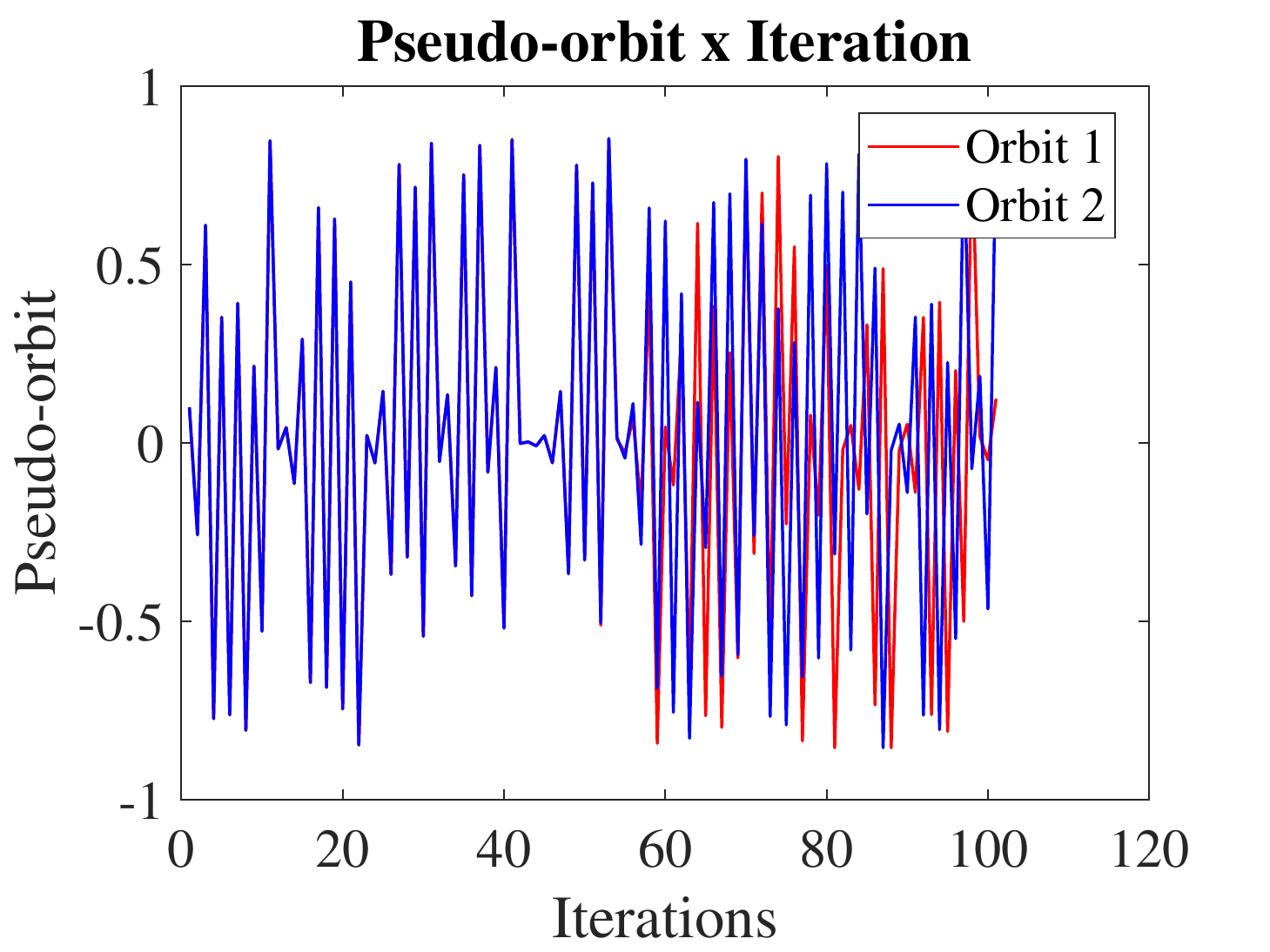}

\caption{Image shows the overlay of two outputs for the cubic map simulated with the same set of initial conditions but in different devices. Since both have somewhat different representation standards, we thereby call each result a pseudo-orbit.}
\label{fig:resultsoverlay}
\end{figure}

The problem of sharing an encrypted file across different devices was studied. According to previous researches \citep{CHAI2017197,Li2017,7932843,pop00001,pop00002,pop00003} and to the best of our knowledge, few concerns were raised about the reliability of the chaotic system simulation. In our study, we have analyzed how chaos-based encryption is applied for information exchange, and developed a method to raise the chance of success of the decryption phase. Obtained results of the simulations were compared iteration by iteration. Figure \ref{fig:resultsoverlay} shows the chaotic output for the $100$ first calculations on each device.

A qualitative analysis to determine the randomness of both $Key$ matrix was applied, based on the normalized Shannon entropy method and calculating the probabilistic distribution of the output vector. Table \ref{tb:entropyshannon} shows that entropy values differs from one another, which also reassure that the two pseudo-orbits are different. 

\begin{table}[!ht]
\centering
\caption{\label{tb:entropyshannon} Entropy values for each device.}
\begin{tabular}{lc}
\hline
\textbf{Environment} & \textbf{Shannon Entropy}\\ \hline
Device 1 & 0.97206928 \\
Device 2 & 0.97283489\\ \hline\\
\end{tabular}
\end{table}

Nevertheless, Table \ref{tb:entropyshannon} shows great values of entropy, matching the expectations for a chaotic output and what is present in the literature.

\begin{figure*}[!htb]
\centering
\begin{tabular}{ccc}
     (a) & (b) & (c) \\
   \includegraphics[width=0.31\textwidth]{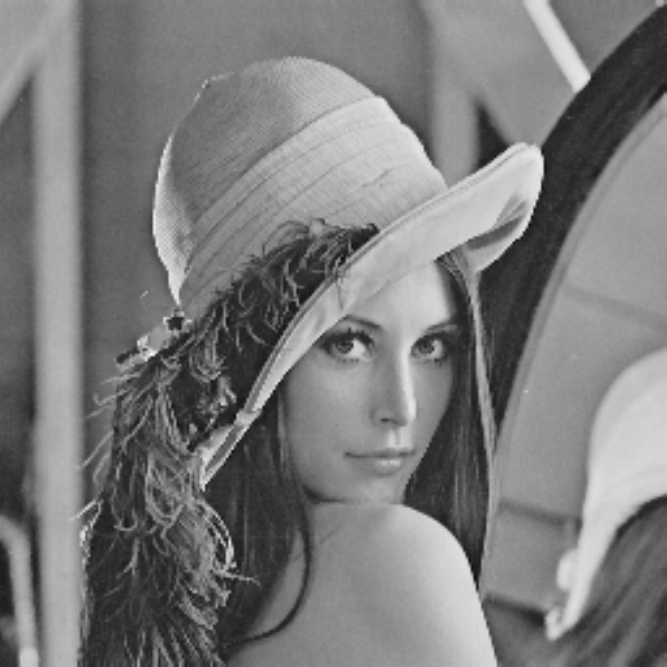} &   
   \includegraphics[width=0.31\textwidth]{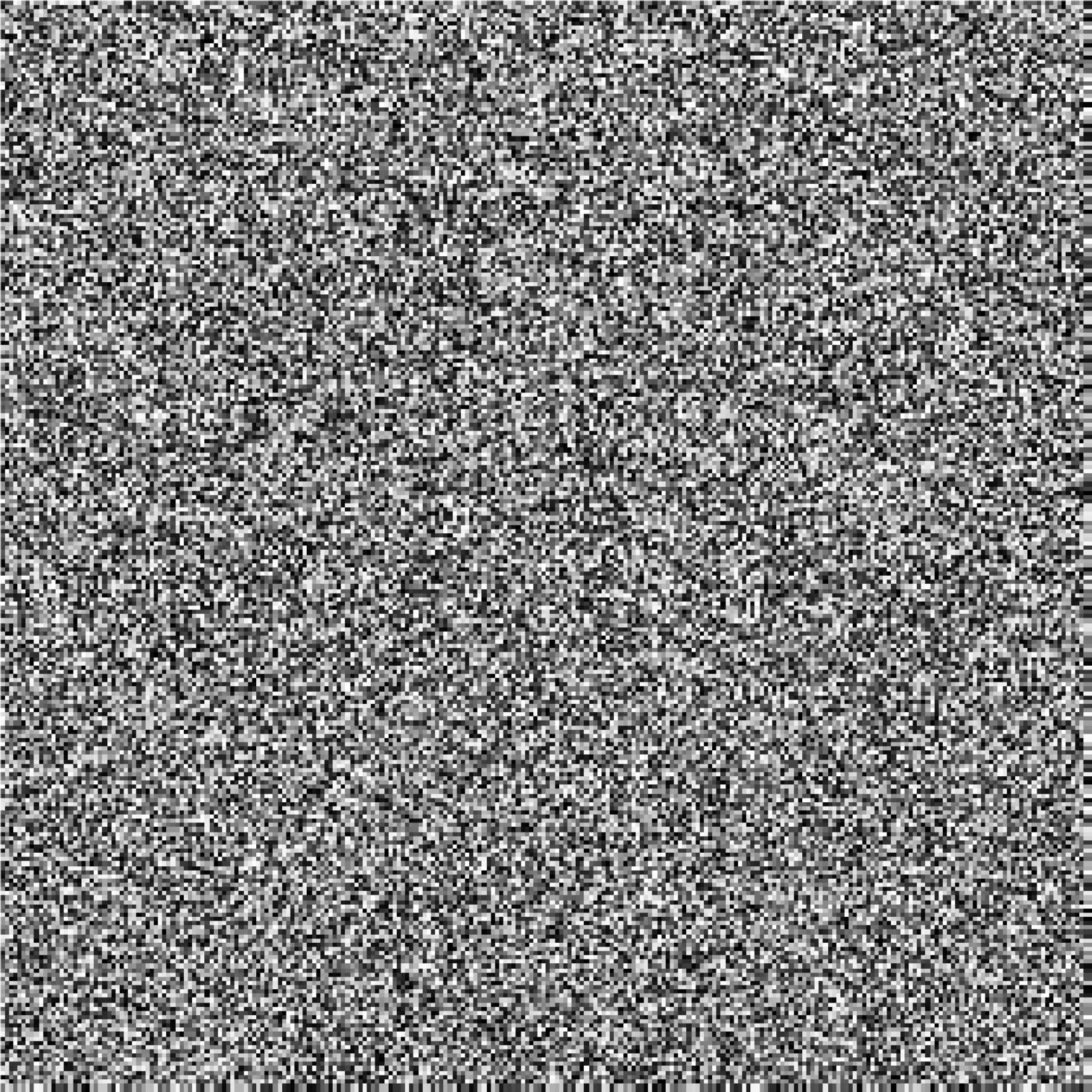} &
    \includegraphics[width=0.31\textwidth]{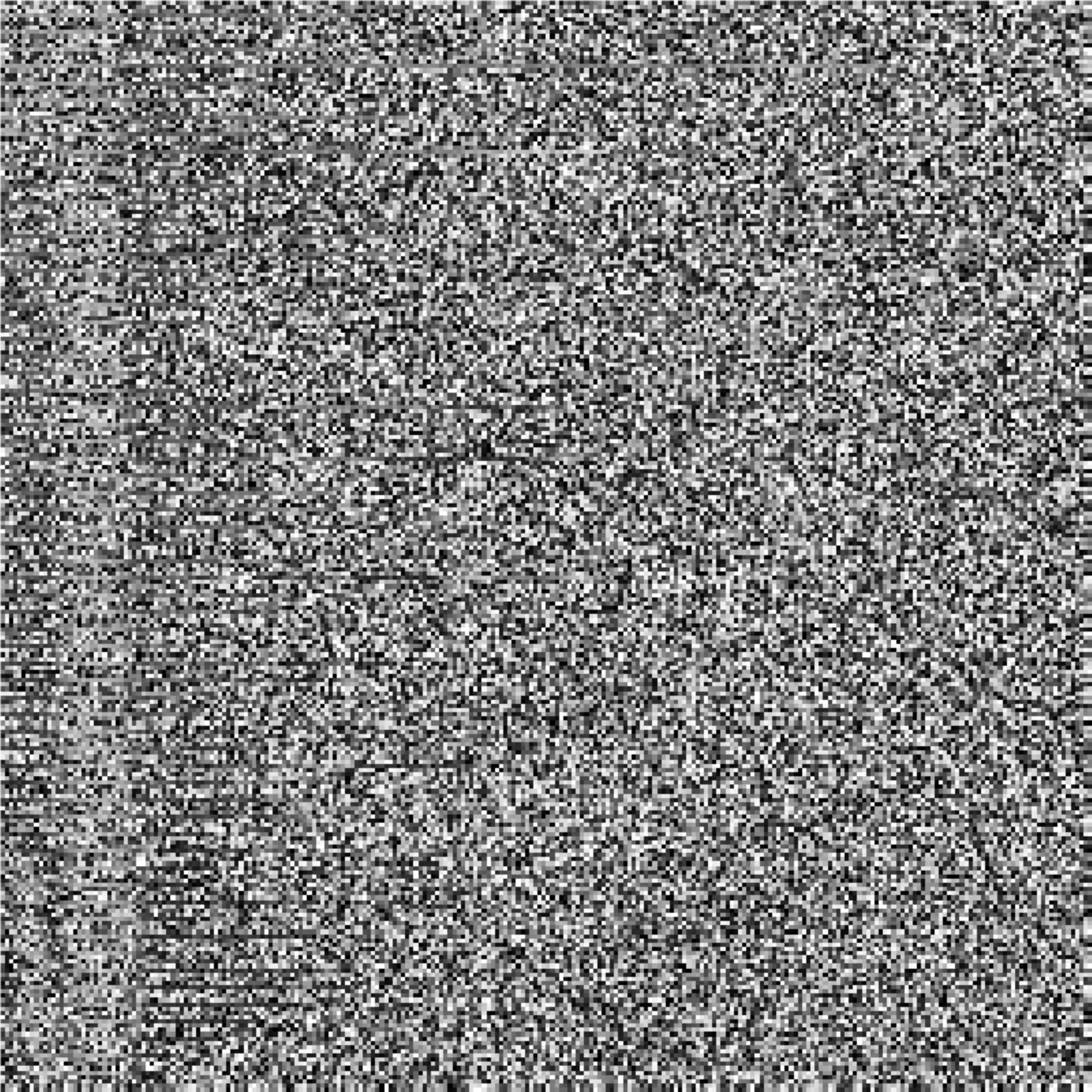}
\end{tabular}
\caption{(a) shows the original image. (b) is the result of the bit-XOR operation between the original image and one pseudo-orbit. (c) is the result of the bit-XOR operation between (b) and the other pseudo-orbit. If both pseudo-orbits were identical, (c) should become the original image, but instead, it's completely unreadable.}
\label{fig:processnormal}
\end{figure*}

After simulating the system, matrix $Key$ from Device 1 was used as input to Equation \ref{eq:encryption}, and operated with the original Figure \ref{fig:processnormal}(a) to generate Figure \ref{fig:processnormal}(b). The latter was then operated with matrix $Key$ from Device 2 and yielded Figure \ref{fig:processnormal}(c). The decryption process failed, as can be seen in Figure \ref{fig:processnormal}(c), as it was expected the original image to be shown, but instead the output remained unreadable.

\begin{figure*}[!htb]
\centering
\begin{tabular}{ccc}
     (a) & (b) & (c) \\
   \includegraphics[width=0.31\textwidth]{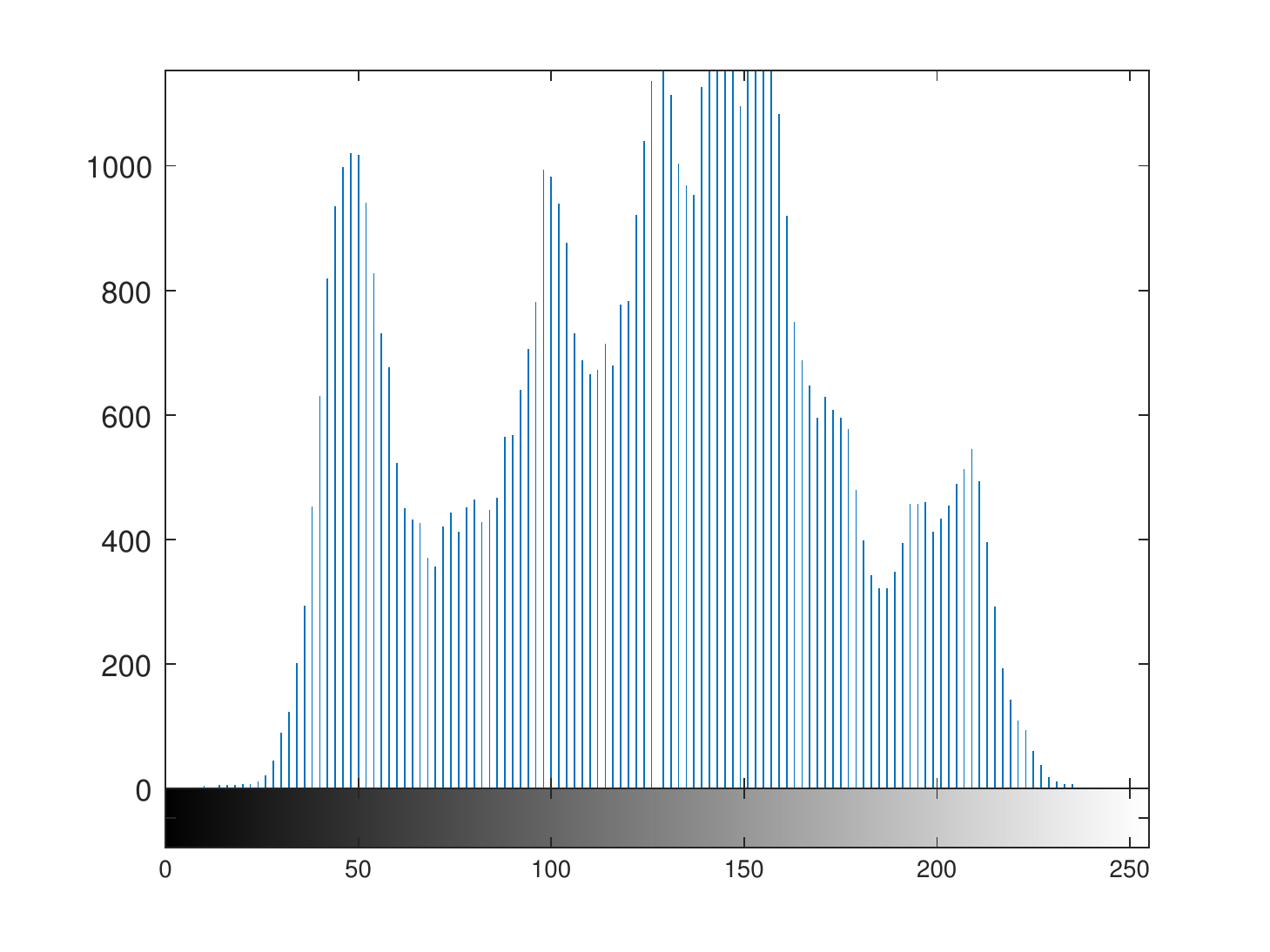} &   
   \includegraphics[width=0.31\textwidth]{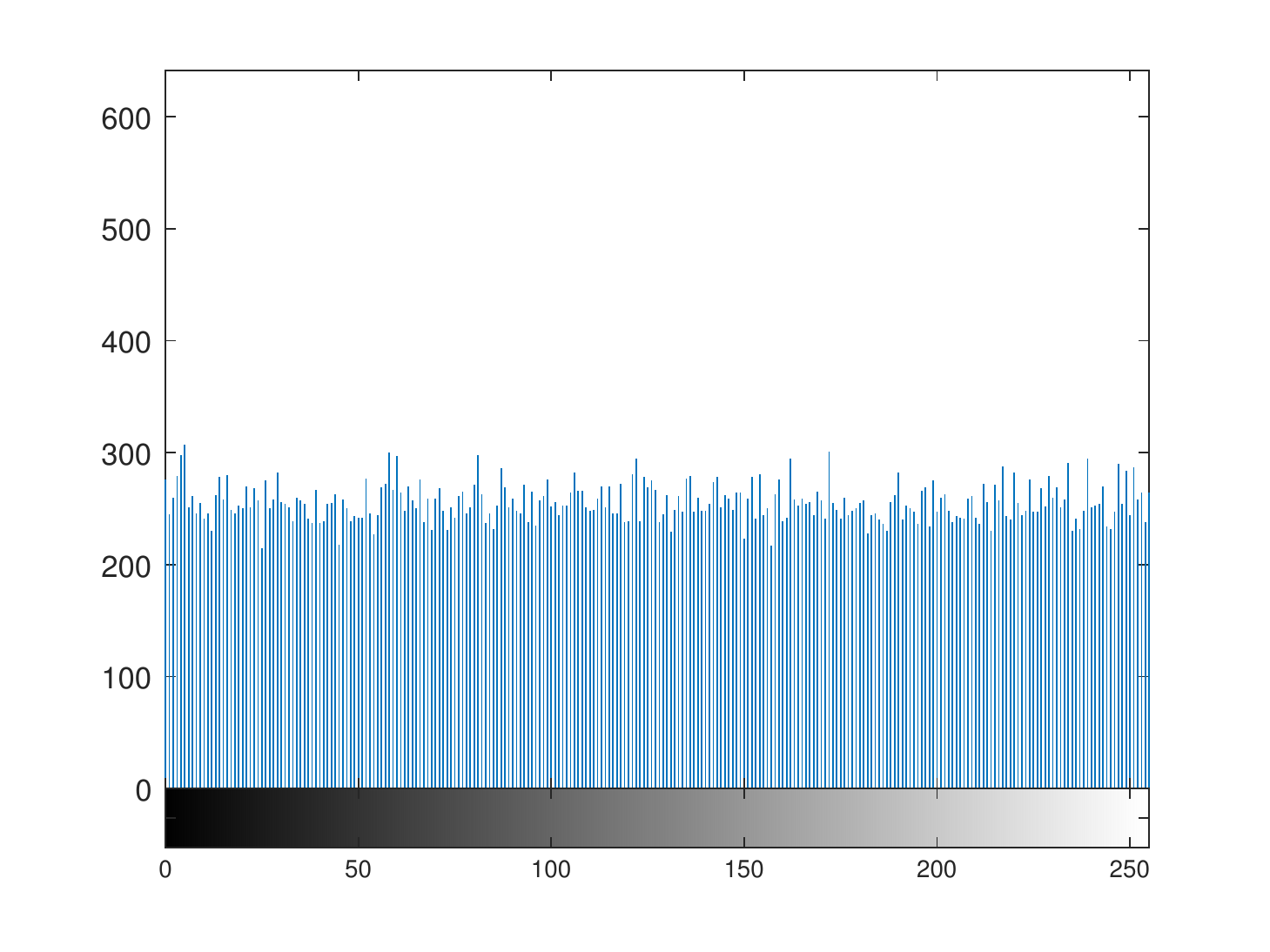} &
    \includegraphics[width=0.31\textwidth]{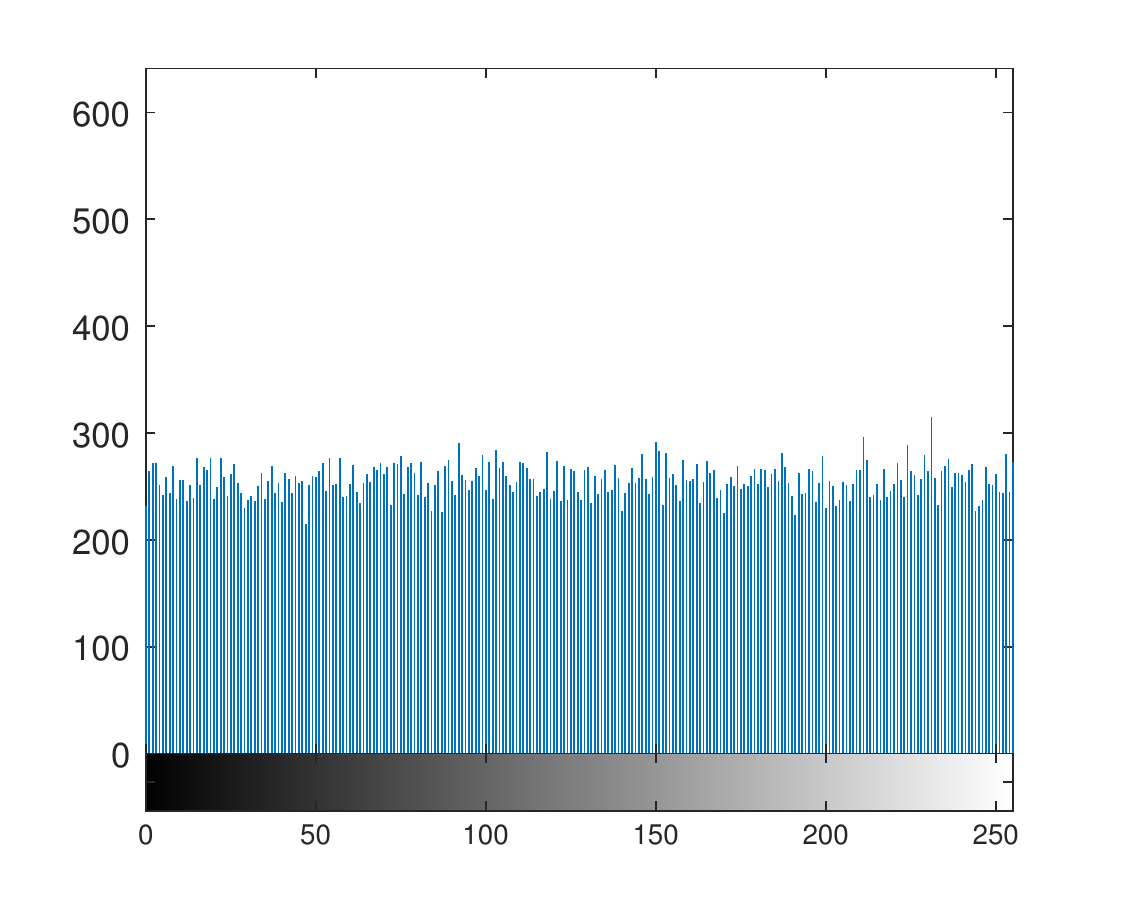}
\end{tabular}
\caption{(a) shows the original image. (b) is the result of the bit-XOR operation between the original image and one pseudo-orbit. (c) is the result of the bit-XOR operation between (b) and the other pseudo-orbit. If both pseudo-orbits were identical, (c) should become the original image, but instead, it's completely unreadable.}
\label{fig:histograms}
\end{figure*}

Figure \ref{fig:histograms}(a) illustrates the histogram of the original image, Figure \ref{fig:histograms}(b) illustrates the histogram of the encrypted image and the Figure \ref{fig:histograms}(c) illustrates the histogram of the decrypted image by means of the gray-scale frequency analysis. It can be seen that the histogram of Figure \ref{fig:histograms}(b) - encrypted - is quite uniform and is significantly different from the histogram of the original image given by Figure \ref{fig:histograms}(a). The more homogeneous (evenly distributed) the histogram is, the more diffuse the image will be. It is observed that the histogram of the original image is not homogeneous, that is, it has different behavior in relation to the encrypted image, whose graph is used to indicate the quality of the cryptography. It is also noted that in Figure \ref{fig:histograms}(a) there are small apertures between the columns, which are filled as soon as the encryption algorithm is executed. 

It can be seen that inherit characteristics of the different processor construction interfered in the process of generating the same key. It was not possible to generate the same temporal series from different devices. The error propagation and limited precision of floating point representation makes such cryptography unreliable for data exchange between different devices, since any minor divergence in the numerical representation could have great impact in system's synthesis, and consequently, key generation. 

\begin{figure*}[!htb]
\centering
\begin{tabular}{ccc}
     (a) & (b) & (c) \\
   \includegraphics[width=0.31\textwidth]{lenafunc.pdf} &   
   \includegraphics[width=0.31\textwidth]{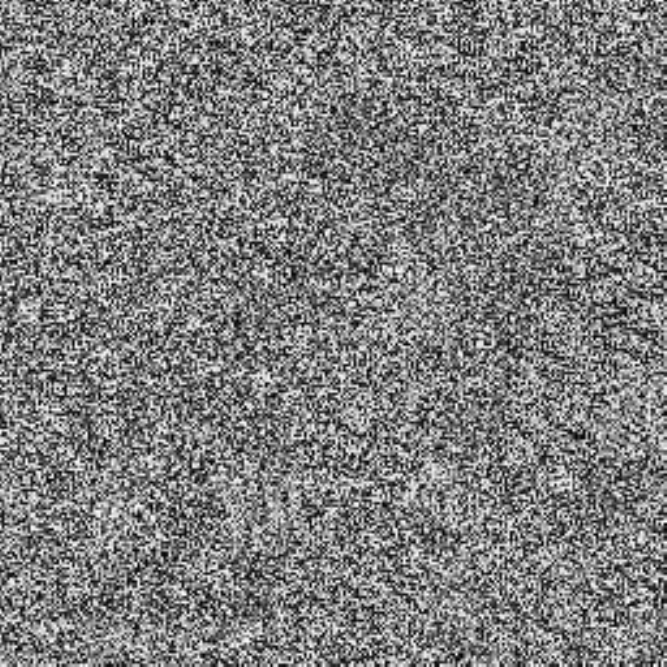} &
    \includegraphics[width=0.31\textwidth]{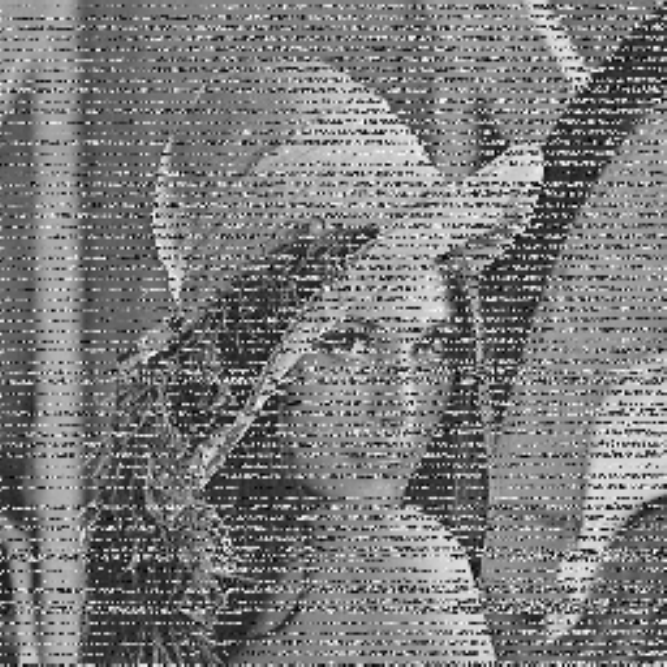}
\end{tabular}
\caption{(a) shows the original image. (b) is the result of the bit-XOR operation between the original image matrix and Device 1 generated key. (c) is the result of the bit-XOR operation between (b) and the key generated by Device 2, after data exchange. If both pseudo-orbits were identical, (c) should become the original image, but instead, it's completely unreadable, as simulation results between Device 1 and 2 were different.}
\label{fig:method}
\end{figure*}

\begin{figure}[!htb]
\centering
\includegraphics[width=7cm]{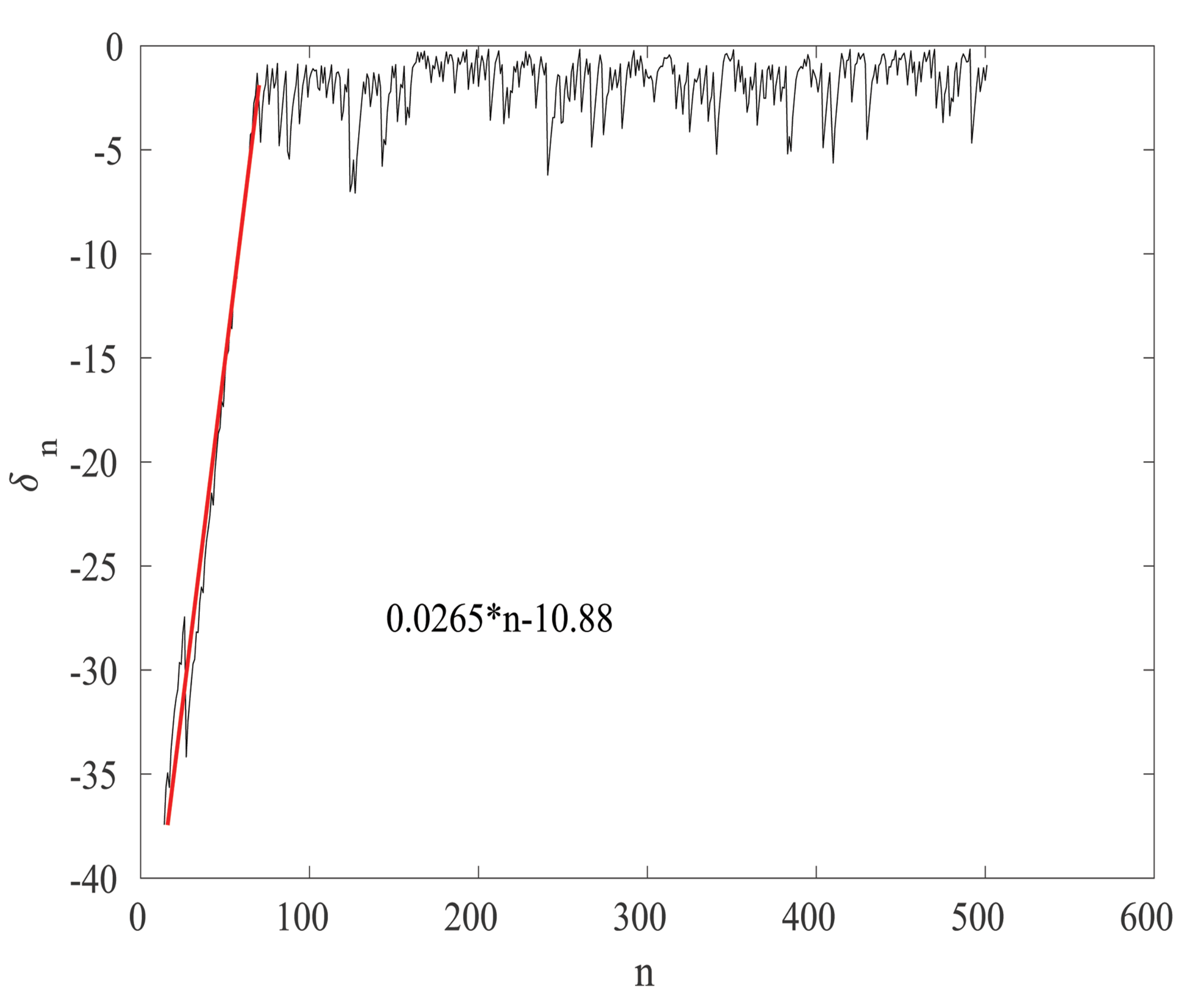}

\caption{Lower bound error ($\delta _{n}$), and calculated maximum Lyapunov exponent from $\delta_{n}$ linear regression. The simulation approaches great values of uncertainty within 70 iterations, as it has greater Lyapunov exponent value.}
\label{fig:llenormal}
\end{figure}

\begin{figure}[htb]
\centering
\includegraphics[width=7cm]{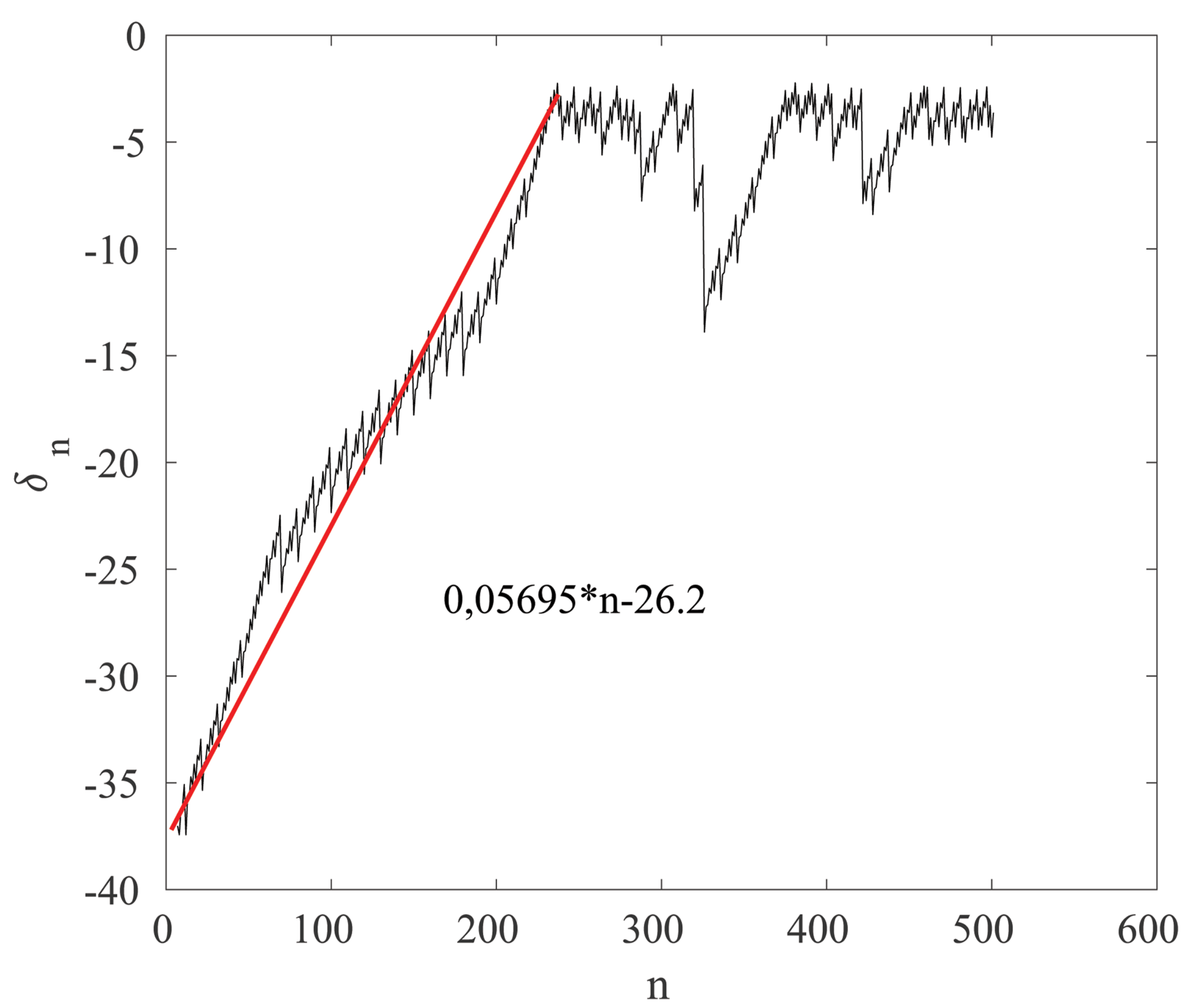}
\caption{Lower bound error ($\delta _{n}$), and calculated maximum Lyapunov exponent from $\delta _{n}$ linear regression. It can be observed that a minor exponent was obtained, and the simulation took more iterations to loose precision.}
\label{fig:llereduction}
\end{figure}

As an attempt to enable some interpretation of the decrypted file, each iteration result was multiplied by $0.89$, and expected to have a reduction of the largest Lyapunov exponent. Such expectation can be confirmed by comparing Figure \ref{fig:llenormal} and Figure \ref{fig:llereduction}, which means that the simulation acquired greater levels of trust for the same number of iterations \cite{llenepomuceno}. As the process gets repeated, now with Lyapunov reduction, a partially readable decryption is possible, as shown in Figure \ref{fig:method}.

%expoente de lyapunov 0.7192

\section{Conclusion}
The problem of sharing an encrypted file across different devices was studied. It has been shown that many existing chaos-based encryption schemes are just special cases of computational arithmetic properties, where non-user accessible conditions of the processor architecture matched. Also, authors observed that even applying techniques to get more reliability over computational simulation, additional treatment was needed in order to perform the decryption process. Although unable to recover the image integrally, the proposed method presents a solution were minor Lyapunov exponents can be reduced to ensure reliability of the simulation. It should be noted that this solution only applies to small values of Lyapunov exponent, and other techniques should be developed to handle bigger values.

\bibliography{referencias}

\end{document}